\documentclass[pra,twocolumn,showpacs]{revtex4}
\usepackage{epsfig}
\usepackage{graphicx}
\usepackage{amsmath}
\usepackage{natbib}
\newcommand{\bn}{\begin{eqnarray}}
\newcommand{\en}{\end{eqnarray}}
\newcommand{\eml}{\end{multline}}
\newcommand{\bml}{\begin{multline}}

\begin{document}

\title {Dynamical spin squeezing: combining fast one-axis twisting and deep two-axis counter-twisting}
 \author{Tom\'{a}\v{s} Opatrn\'{y}}
 \affiliation{Optics Department, Faculty of Science, Palack\'{y} University, 17. Listopadu 12,
 77146 Olomouc, Czech Republic}

\date{\today }
\begin{abstract}
Based on the recent twisting-tensor approach [T. Opatrn\'{y}, ArXiv:1408.3265 (2014)], a specific scenario for fast and deep spin squeezing is proposed. Initially the state is subjected to one-axis twisting under optimum orientation, enabling the maximum squeezing rate allowed by the system nonlinearity. Later on, when for highly squeezed states the one-axis twisting deforms the uncertainty ellipse and deteriorates the squeezing properties, the process is switched to an effective two-axis counter-twisting by a sequence of $\pi/2$ pulses. The squeezing rate then slows to 2/3 of the maximum value, but the process can continue for longer to achieve a very high degree of squeezing.
\end{abstract}
%
%

\maketitle

\section{Introduction}
Recent experimental success in producing spin squeezing in atomic samples with collisional nonlinearity \cite{Esteve2008,Gross2010,Riedel} stimulated search for optimum squeezing strategy under the condition of given system nonlinearity. Various proposals of ``shortcuts to adiabaticity'' \cite{Diaz2012,Yuste,Calarco} and optimized pulse sequences \cite{Liu2011,Shen2013,Zhang2014} have been studied. Starting with the pioneering work by Kitagawa and Ueda \cite{Kitagawa}, two basic squeezing schemes have been identified, the so called one-axis twisting (OAT, Hamiltonian $\sim J_z^2$), and two-axis counter-twisting (TACT, Hamiltonian $\sim J_x^2-J_y^2$). It has generally been recognized that the TACT is more efficient in generating highly squeezed states, although it may be much more challenging to realize  experimentally. Therefore, some of the schemes focused at producing effective TACT from a $\sim J_z^2$ Hamiltonian by displacing the state back and forth on the Poincar\'{e} sphere by suitable $J_y$ or $J_x$-pulses \cite{Liu2011,Zhang2014}.

Recently, it has been shown that all Hamiltonians quadratic in $J$ can be treated in a unified twisting-tensor formalism, and that the maximum squeezing rate is given by the difference of the maximum and minimum eigenvalues of the twisting tensor \cite{TO2014}. Thus, although  by applying a sequence of rotations one can change the Hamiltonian from the OAT character to the TACT, one does not increase the maximum squeezing rate. In fact, the emulated TACT of \cite{Liu2011,Zhang2014} works at most at 2/3 of the maximum rate achieved by the OAT at initial stages. Here we propose a scheme combining the advantages of these two approaches such that at the beginning the state is squeezed by the maximum possible rate, and later, when the OAT becomes less efficient, the process is switched to the effective TACT scheme. Although the squeezing rate slows down, ultimately the state can be squeezed much deeper than when using the OAT model only.

\section{System description}
The system is composed of two bosonic modes described by annihilation operators $a$ and $b$ satisfying the commutation relation $[a,a^{\dag}] = [b,b^{\dag}] =1$. If the processes conserve the total particle number $N=a^{\dag}a + b^{\dag}b$, it is convenient to introduce orbital momentum-like operators $J_{x,y,z}$ defined as 
\begin{eqnarray}
J_x &=& \frac{1}{2}(a^{\dag}b+ab^{\dag}), \\
J_y &=& \frac{1}{2i}(a^{\dag}b-ab^{\dag}), \\
J_z &=& \frac{1}{2}(a^{\dag}a-b^{\dag}b), 
\label{eqabN}
\end{eqnarray}
satisfying $[J_x,J_y]=iJ_z$,  $[J_y,J_z]=iJ_x$, and $[J_z,J_x]=iJ_y$.
Let the Hamiltonian be composed of $a$, $b$, $a^{\dag}$, and  $b^{\dag}$ such that in each term the same number of creation and annihilation operators occurs (total number of particles is conserved), and the highest power of each operator is 2. In this case the Hamiltonian can be written as
\begin{eqnarray}
H = \omega_k J_k +\chi_{kl} J_k J_l + f(N), 
\label{Ham}
\end{eqnarray}
where  ${\bf \omega} = (\omega_x, \omega_y, \omega_z)$ transforms as a vector and $\chi_{kl}=\chi_{lk}$ transforms as a tensor under O(3) rotations (we call $\chi$ the ``twisting tensor'' \cite{TO2014}).
In Eq. (\ref{Ham}), Einstein summation rule is applied, and $f(N)$ is a linear or quadratic function of the total particle number,  generating an unimportant overall phase. As discussed in \cite{TO2014}, it is convenient to select such a coordinate system in which the twisting tensor is diagonal. Since $J_x^2+J_y^2+J_z^2=\frac{N}{2}(\frac{N}{2}+1)$, addition of an arbitrary multiple of unit matrix to $\chi$ can be absorbed in the unimportant term $f(N)$. Therefore, one can always choose it such that the middle eigenvalue of $\chi$ is zero.

There are two special cases of the twisting tensor. First, the tensor is degenerate with two eigenvalues equal to each other. Setting these eigenvalues to zero and denoting the nonzero eigenvalue $\chi$ (it should be clear from the context when $\chi$ denotes the whole tensor and when its components), the twisting tensor is 
\begin{eqnarray}
\chi^{\rm (OAT)} = \left( 
\begin{array}{ccc}
0 & 0 & 0  \\
0 & 0 & 0 \\
0 & 0 & \chi
\end{array}
\right) ,
\label{chi1ax}
\end{eqnarray}
corresponding to OAT. 

The second special case is if the  central eigenvalue is exactly in the middle of the two remaining ones. Choosing the central (zero) eigenvalue to correspond to the $z$-direction, the twisting tensor has the form 
\begin{eqnarray}
\chi^{\rm (TACT)} = \left( 
\begin{array}{ccc}
\chi & 0 & 0  \\
0 & -\chi & 0 \\
0 & 0 & 0
\end{array}
\right) ,
\label{chi2ax}
\end{eqnarray}
corresponding to TACT.

\begin{figure}
\centerline{\epsfig{file=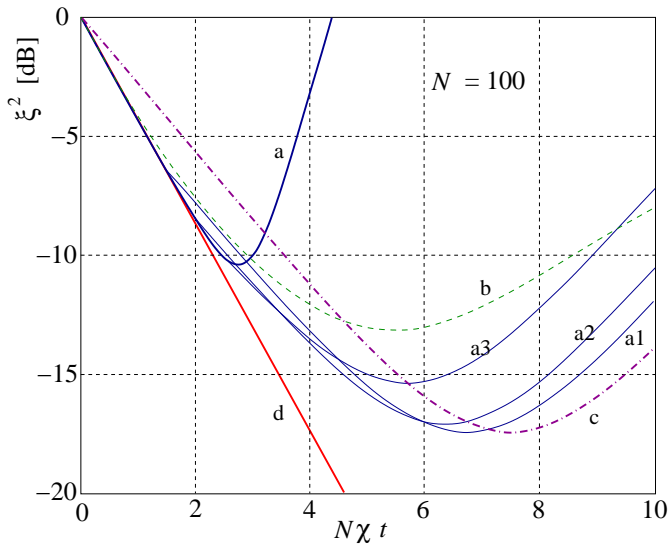,scale=0.38}}
\caption{\label{fcomp1} (Color online.) Time evolution of the squeezing parameter $\xi^2$ with $N=100$ particles. Line a: OAT with optimum orientation of the state; line b: OAT without rotating the state; line c: TACT emulation; line d: asymptotics of the maximum squeezing rate for $N\to \infty$, Eq. (\ref{asymptot}). Lines a1---a3 correspond to switching from the OAT to the emulated TACT at various times: $N\chi t_{\rm switch} = 1.5$ (a1); 2.0 (a2);  2.5 (a3). The length of one cycle of the TACT emulation is $N\chi t_{\rm cycle} = 0.04$.}
\end{figure}

\begin{figure}[t]
\centerline{\epsfig{file=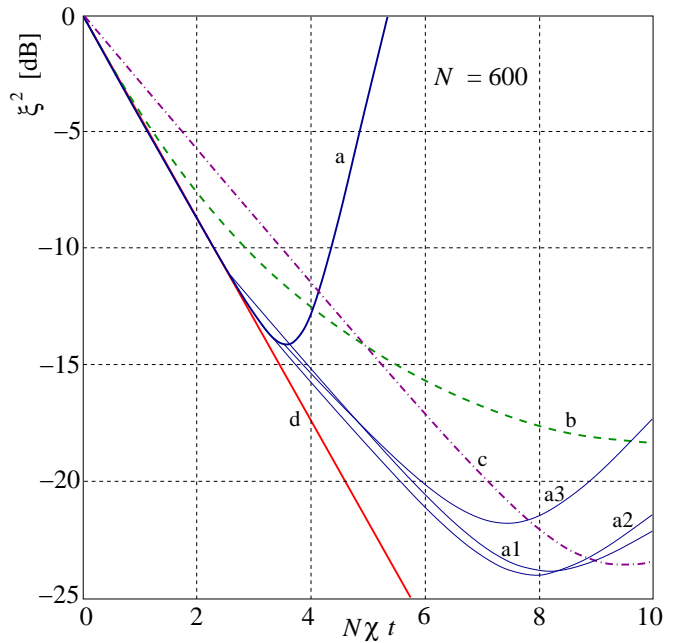,scale=0.38}}
\caption{\label{fcomp2} (Color online.) Same as Fig. \ref{fcomp1} but with larger particle number, $N=600$.  Lines a1---a3 correspond to the following switching times: $N\chi t_{\rm switch} = 2.5$ (a1); 3.0 (a2); 3.5 (a3). }
\end{figure}

\section{Squeezing rate }
As shown in \cite{TO2014}, for nearly Gaussian states (weakly squeezed spin coherent states) the maximum squeezing rate only depends on the difference between the maximum and the minimum eigenvalues of the twisting tensor, in particular
\begin{eqnarray}
{\rm max} (Q) = N \left( \chi^{\rm (max)} - \chi^{\rm (min)}\right) ,
\label{Qmax}
\end{eqnarray}
where $Q$ is defined by
\begin{eqnarray}
 \frac{d \xi^2}{dt} = -Q \xi^2 .
 \label{xiQ}
\end{eqnarray}
Here the squeezing parameter is defined as $\xi^2  \equiv  4V_-/N$, where $V_-$ is the smallest eigenvalue of the variance matrix $V$, which for a state centered at the equator of the Poincar\'{e} sphere with $\langle J_y\rangle = \langle J_z\rangle = 0$ is
\begin{eqnarray}
V = \left( 
\begin{array}{cc}
V_{yy} & V_{yz}  \\
V_{zy} & V_{zz} 
\end{array}
\right) ,
\label{varV}
\end{eqnarray}
with
\begin{eqnarray}
 V_{yy} &=& \langle J_y^2 \rangle, \\ 
 V_{zz} &=& \langle J_z^2 \rangle, \\ 
 V_{yz} &=& \frac{1}{2}\langle J_y J_z + J_z J_y \rangle.
\end{eqnarray}
The squeezing rate of Eq. (\ref{xiQ}) depends on the position of the state on the Poincar\'{e} sphere and on the orientation of the squeezing ellipse. The maximum of Eq. (\ref{Qmax}) is achieved for states centered at the direction of the main axis corresponding to the middle eigenvalue of $\chi$, for which orientation of the squeezing ellipse is $\pi/4$ with respect to the direction of the pole of the maximum eigenvalue (for details see \cite{TO2014}). If $\chi$ is degenerate (i.e., OAT case) with the nonzero eigenvalue in the $z$-direction as in Eq. (\ref{chi1ax}), the fastest squeezing is achieved for states along the equator with $\langle J_z \rangle=0$, the optimum orientation of the uncertainty ellipse being $\pi/4$ with respect to the meridian.

\begin{figure}[t]
\centerline{\epsfig{file=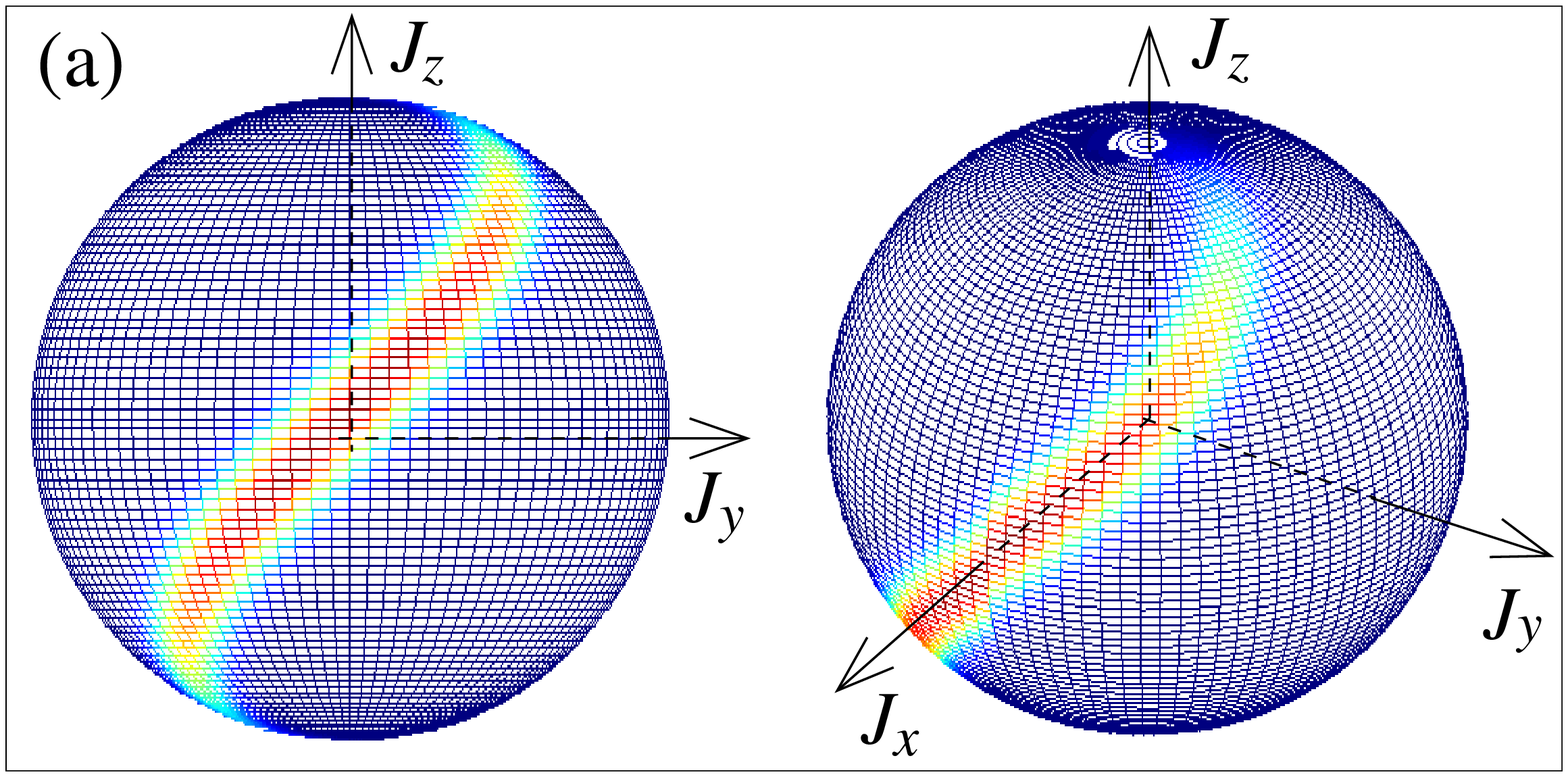,scale=0.28}}
\centerline{\epsfig{file=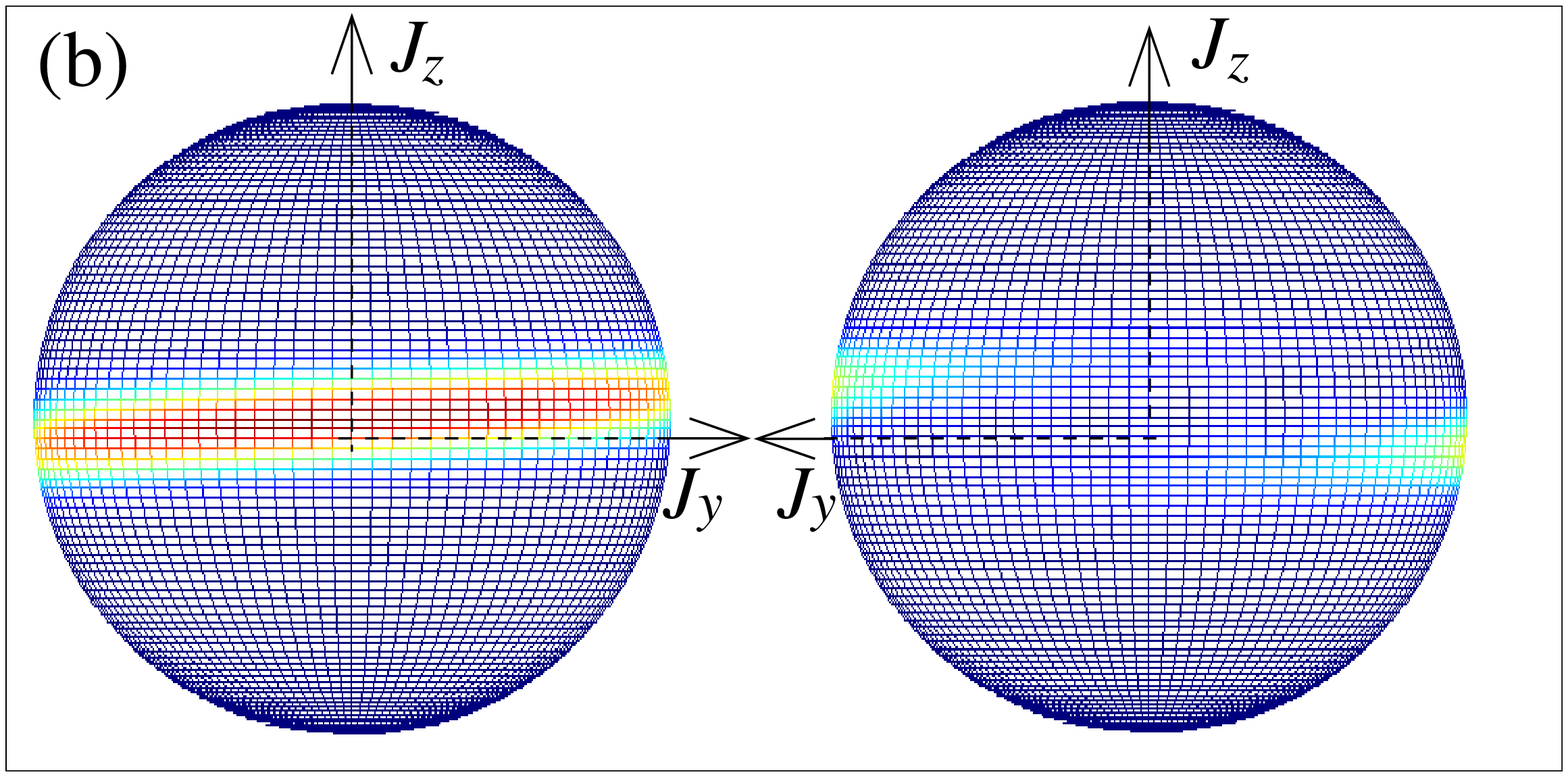,scale=0.28}}
\centerline{\epsfig{file=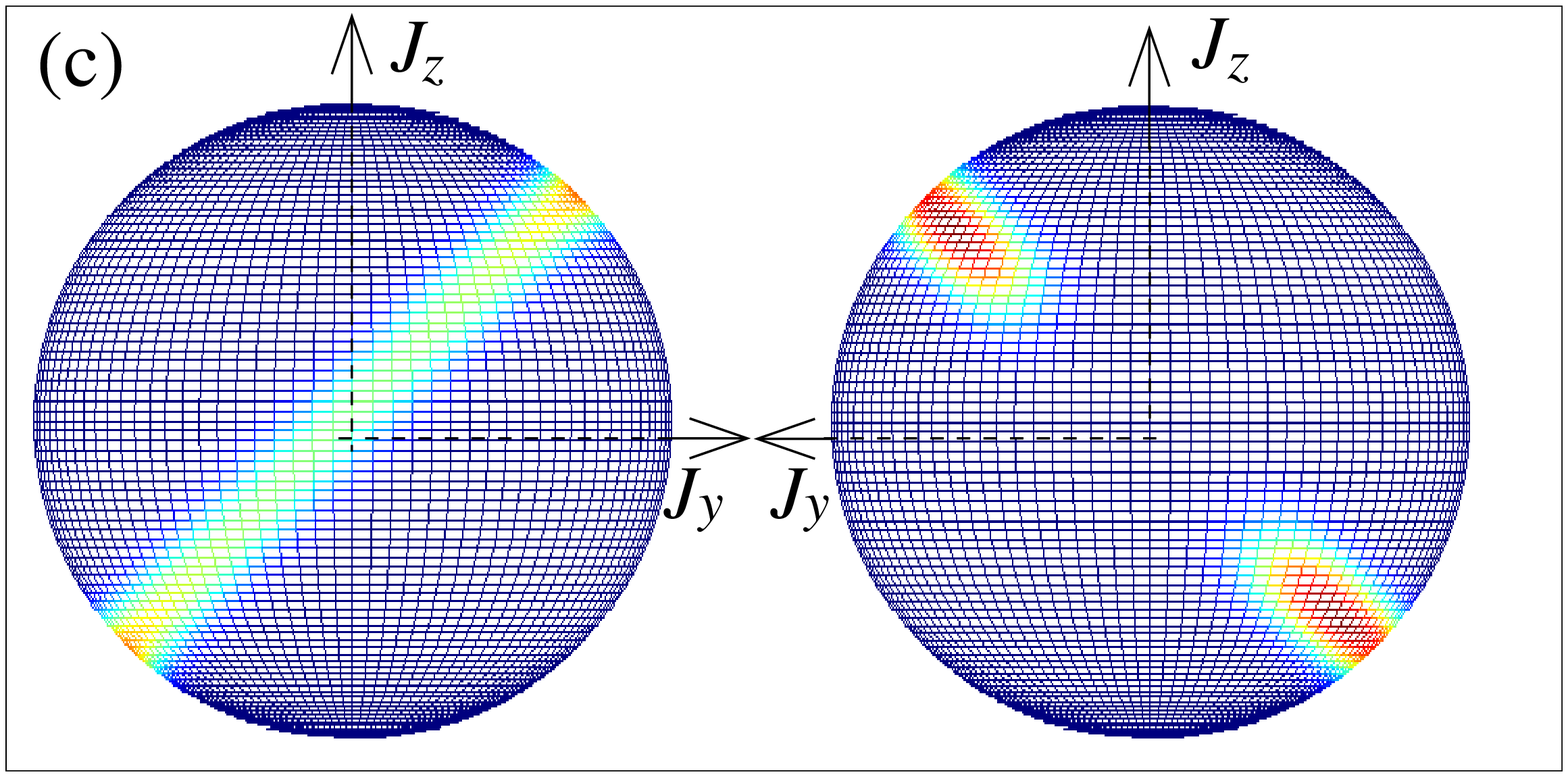,scale=0.28}}
\caption{\label{fspheres} (Color online.) Q-function of the state deformed in various processes, with $N=100$. (a) OAT with optimized rotation, curve `a' in Fig. \ref{fcomp1} at $N\chi t = 4.5$; (b) OAT without rotation, curve `b' in  Fig. \ref{fcomp1} extended to $N\chi t = 16$; (c) emulated TACT, curve `c' in  Fig. \ref{fcomp1} at $N\chi t = 10$. }
\end{figure}

\section{Fast one-axis twisting}
Thus, to squeeze an initially spin coherent state the fastest way, one has to place it on the equator of the Poincar\'{e} sphere and keep its optimum $\pi/4$ orientation. If the state is initially located, e.g., in the $J_x$ direction, then, as shown in \cite{TO2014}, the optimum orientation is kept fixed by rotating the state around $J_x$ with the frequency 
\begin{eqnarray}
 \omega_x = \frac{N\chi}{2} .
\end{eqnarray}
This dynamics corresponds to the early squeezing around an unstable point studied in  \cite{Diaz2012a} with the Hamiltonian $H=\chi \left( \frac{N}{\Lambda} J_x + J_z^2 \right)$ with $\Lambda =2$.
In the limit of $N\to \infty$ the squeezing parameter would decrease exponentially as 
\begin{eqnarray}
 \xi^2 = \exp \left( - N\chi t \right) 
 \label{asymptot}
\end{eqnarray}
(see Figs. \ref{fcomp1} and \ref{fcomp2}, line `d').
However, for finite $N$ the state starts deviating from Gaussian, after some time its shape becomes $S$-deformed and the squeezing deteriorates. The Q-function of a strongly deformed state that underwent this process is in Fig. \ref{fspheres}a and the evolution of $\xi^2$ is in Figs. \ref{fcomp1} and  \ref{fcomp2}, line `a'. 

Note that if we do not rotate the state to keep the optimal orientation, the squeezing parameter decreases slower than exponentially, but eventually can reach deeper values (see Figs. \ref{fcomp1} and  \ref{fcomp2}, line `b'). For large $t$  the state is deformed as the Q-function winds around the Poincar\'{e} sphere in a spiral (see Fig. \ref{fspheres}b).

\begin{figure}[t]
\centerline{\epsfig{file=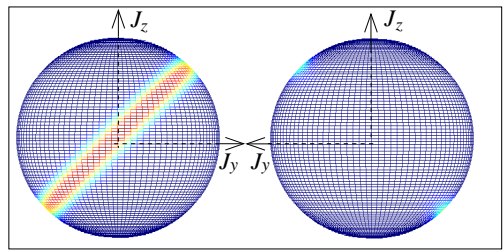,scale=0.28}}
\caption{\label{fspheres4} (Color online.) Q-function of the state with the smallest $\xi^2$ achieved by the combined method with $N=100$, curve `a1' in Fig. \ref{fcomp1} at $N\chi t = 6.7$. The same shape corresponds also to the Q-function of the TACT, curve `c' at $N\chi t = 7.5$. }
\end{figure}

\section{Deep two-axis counter-twisting}
In \cite{Liu2011} and \cite{Zhang2014} a scheme has been proposed to generate an effective TACT by switching the position of the state between two locations, one on the equator and the other on the pole of the Poincar\'{e} sphere. The argument was based on the Baker-Campbell-Hausdorff expansion \cite{Liu2011}, or the  Trotter-Suzuki expansion \cite{Zhang2014} of operator exponentials. The simplest scheme of \cite{Liu2011} works as follows. The state originally at the equator of the Poincar\'{e} sphere centered at $\langle J_y\rangle = \langle J_z\rangle =0$ evolves first under the Hamiltonian $\chi J_z^2$ for 2/3 of the cycle period $t_{\rm cycle}$, then a $\pi/2$ pulse of $J_y$ moves it to the pole where it evolves for  1/3 of the cycle period under $\chi J_z^2$, and finally another $\pi/2$ pulse of $-J_y$ brings it back to the original location. Thus, after one cycle the state is changed by the operator
\begin{eqnarray}
 U&=& \exp\left( -i \frac{\pi}{2} J_y  \right)  \exp\left( -i \frac{1}{3}\chi J_z^2  t_{\rm cycle} \right) 
\nonumber \\
& & \times
\exp\left( i \frac{\pi}{2} J_y  \right) \exp\left( -i \frac{2}{3}\chi J_z^2  t_{\rm cycle} \right) .
\end{eqnarray}

The understanding of the schemes is straightforward in terms of the twisting tensor. Rather than moving the state, one can describe the situation in a coordinate system in which the location of the state is fixed and the twisting tensor is switched between two different orientations. In one orientation, the main axis with nonzero $\chi$ is in the $J_z$ direction, in the other orientation it is in the $J_x$ direction. The first case lasts for $2/3$ of the period, the second for $1/3$ of the period. Thus, in average the state is subjected to an effective  twisting tensor with components $(2/3)\chi$ along $J_z$,  $(1/3)\chi$ along $J_x$, and 0 along $J_y$. Since the middle eigenvalue is exactly between the two other eigenvalues, the resulting process is two-axis counter-twisting for which the twisting tensor can be written (after shifting the central eigenvalue to zero) as
\begin{eqnarray}
\chi^{\rm (TACT\ eff)} = \frac{1}{3}\left( 
\begin{array}{ccc}
-\chi & 0 & 0  \\
0 & 0 & 0 \\
0 & 0 & \chi
\end{array}
\right) .
\label{chi2axeff}
\end{eqnarray}
Since the difference between the largest and smallest eigenvalues is $(2/3)\chi$, the maximum squeezing rate is 2/3 of that of the OAT scheme. On the other hand, in the TACT scheme the state suffers much less from the shape deformations and the squeezing process can last longer. Thus, ultimately much deeper squeezing can be reached. This can be seen in Figs. \ref{fcomp1} and  \ref{fcomp2}, line `c'. For longer times the state is also deformed and $\xi^2$ increases, the Q-function being torn apart into two counter-propagating segments (see Fig. \ref{fspheres}c).

\section{Combined scheme}
The goal is to combine advantages of the two schemes such that at the beginning the state is squeezed fast under the OAT and then at time $t_{\rm switch}$ the scheme is switched to the slower TACT to reach deeper squeezing values. The results for two different particle numbers are in Figs. \ref{fcomp1} and  \ref{fcomp2}, lines `a1'--`a3'. 

In our examples, for $N=100$, the best achieved squeezing was $-17.5$ dB. In the emulated TACT scheme it was achieved at $N\chi t = 7.4$, and in the combined scheme with $N\chi t_{\rm switch} = 1.5$ the same amount of squeezing was achieved at  $N\chi t = 6.7$. In both cases the time of one cycle in the TACT emulation was $N\chi t_{\rm cycle} = 0.04$.

For $N=600$, the best achieved squeezing in the emulated TACT  was $-23.6$ dB at  $N\chi t = 9.45$. In the combined scheme with $N\chi t_{\rm switch} = 3.0$ the best achieved squeezing  was $-24.0$ dB at  $N\chi t = 7.96$.  

As can be seen, in the combined scheme deep squeezing is achieved earlier than in the emulated TACT. In our example with  $N=600$ the combined scheme also achieved deeper squeezing. This is caused mostly by the discrete character of the TACT emulation: increasing the number of cycles by decreasing $t_{\rm cycle}$ leads to even deeper squeezing, the optimum values approaching each other in the two methods. For example, shortening the cycle time by a factor of 10, $N\chi t_{\rm cycle} = 0.004$, leads to $-25.17$ dB squeezing in both methods, in the emulated TACT reached at $N\chi t = 10.14$ and in the combined scheme  with $N\chi t_{\rm switch} = 2.0$ at $N\chi t = 9.14$.
 
\section{Conclusion}
The understanding of the  effective TACT scheme can be based on time averaging of the twisting tensor. One can also understand its function as squeezing the state located at the equator for 2/3 of the time, and undoing the $S$-shape deformation at the pole for 1/3 of the time. 
The OAT process generates fourth and higher moments of $J$ differing more from the moments of Gaussian states than moments generated by the TACT process. These changes are partially compensated when the state is relocated to the pole of the Poincar\'{e} sphere.

The combined scheme proposed here takes the advantage from the fact that at the beginning when the state does not deviate far from the equatorial area the $S$-shape deformation is not significant and the state can be squeezed at the maximum possible rate. 
Choice of the switching time $t_{\rm switch}$ and of the TACT-emulation cycle $t_{\rm cycle}$   is matter of optimization dependent on the particular physical system. Since various sources of losses and decoherence cause noises increasing with time, it may not be possible to reach the minimum squeezing generated by pure TACT. In practical situations one would also have to take into account finite amplitude of the fields generating the $\pi/2$ pulses and thus finite duration of the TACT emulation cycle. As a result, a reasonable compromise putting more stress to faster squeezing may be required. 

\acknowledgments
Stimulating discussions with K. M{\o}lmer are acknowledged.
This work was supported by grant IGA PrF 2014008.

\end{document}